\documentclass[a4paper,twoside,twocolumn]{article}
\pdfoutput=1

\newif\iftr 
\trtrue     

\usepackage{ifpdf}  
\ifpdf
    \usepackage[pdftex]{graphicx}
    \usepackage[colorlinks,linkcolor=blue,citecolor=blue,urlcolor=blue]{hyperref}
    \DeclareGraphicsExtensions{.pdf}
\else
    \usepackage{graphicx}
    \usepackage[hypertex]{hyperref}  
\fi

\usepackage{todonotes} 

\usepackage{
url       %
,fancyhdr 
,lastpage 
,enumerate 
,amsmath  
,amsfonts 
,amssymb  
, enumitem 
}
\usepackage[hang]{footmisc} 
\usepackage[noindent, arraystretch, fullpage]{setlengths} 
\usepackage{
own         
}

\usepackage{amsthm}
\newtheorem{assumption}{Assumption}
\newtheorem{requirement}{Requirement}
\newtheorem*{test*}{Test}

\graphicspath{{images/}}

\newcommand*{\metaauthori}{Bob Briscoe}
\newcommand*{\metaauthorii}{Koen De Schepper}
\newcommand*{\metashorttitle}{Scalable Congestion Control Tensions}
\newcommand*{\metatitle}{{\large{Discussion paper}}\\Resolving Tensions between\\Congestion Control Scaling Requirements}
\newcommand*{\metano}{TR-CS-2016-001}
\newcommand*{\metakeywords}{Data Communications, Networks, Internet,
Control, Performance, Latency, Responsiveness, Dynamics, Algorithm, Standards}

\newcommand*{\metamaili}{\href{mailto:research@bobbriscoe.net}{research@bobbriscoe.net}}
\newcommand*{\metamailii}{\href{mailto:koen.de_schepper@nokia.com}{koen.de\_schepper@nokia.com}}
\newcommand*{\metaaddress}{}

\newcommand*{\metaversion}{01}
\newcommand*{\metadate}{11 Jul 2017}

\hypersetup{                       
     pdfauthor = {\metaauthori and \metaauthorii},
     pdftitle = {\metashorttitle},
     pdfsubject = {},
     pdfkeywords = {\metakeywords}
}%

\title{\metatitle}%
\author{\metaauthori%
\thanks{\metamaili, %
\metaaddress}%
\and \metaauthorii%
\thanks{\metamailii}%
}
\date{\metadate}%

\fancypagestyle{first}{%
\fancyhead[LO,RE]{}%
\fancyhead[LE,RO]{}%
\fancyhead[C]{}%
}%

\begin{document}
\bibliographystyle{alpha}%


\maketitle%
\thispagestyle{first}

\begin{abstract}

Low Latency, Low Loss Scalable throughput (L4S) is being proposed as the new default Internet service. L4S can be considered as an `incrementally deployable clean-slate' for new Internet flow-rate control mechanisms. Because, for a brief period, researchers are free to develop host and network mechanisms in tandem, somewhat unconstrained by any pre-existing legacy.

Scaling requirements represent the main constraints on a clean-slate design space. This document confines its scope to the steady state. It aims to resolve the tensions between a number of apparently conflicting scalability requirements for L4S congestion controllers. It has been produced to inform and provide structure to the debate as researchers work towards pre-standardization consensus on this issue.

This work is important, because clean-slate opportunities like this arise only rarely and will only be available briefly---for roughly one year. The decisions we make now will tend to dirty the slate again, probably for many decades.
\end{abstract}

\section{Introduction}\label{sec:ccditr_intro}

A new Internet service has been proposed called L4S, for Low Latency, Low Loss Scalable throughput. It enables so-called `Scalable' congestion controls to keep queuing delay and congestion loss to extremely low levels. But they can still share Internet capacity with existing traffic, while remaining isolated from its highly variable queuing delay and loss. The best background reference on L4S, for the present document is~\cite{DeSchepper15b:DCttH_TR}\footnote{This reference is a little dated. A more up to date paper is under submission. \cite{Briscoe16a:l4s-arch_ID} explains how the parts of L4S fit together and collects together all the references to the details of each part. \url{https://riteproject.eu/dctth/} collects together links to L4S materials.}

For a brief period of a year or so, L4S provides what has been called an `incrementally deployable clean-slate' for new flow-rate control mechanisms. During this brief period, researchers have the freedom to develop congestion controls (CCs) in tandem with queue management mechanisms, because there is no legacy for the L4S traffic class, either on hosts or in the network. 

The aim of this paper is to articulate the tensions between a number of conflicting scaling requirements. The scope is limited to the steady-state. Various ways to resolve these tensions are given, including consideration of whether each requirement is best resolved in the network or on hosts. The idea is to determine the design space that flow-rate control mechanisms are confined to, because scaling requirements are the main constraints on a clean-slate design space.

Realistically, L4S is not truly a clean slate; it is a `slightly-dirty slate', because it is built within the Internet architecture, which imposes a number of additional constraints. Some of these are documented explicitly as assumptions. But many implicit assumptions remain hidden, by definition!

This work is important, because clean-slate opportunities like this arise only rarely and will only be open briefly. The period of research freedom will end as experimental standards start to be approved for the network mechanisms (perhaps late 2017 or early 2018). Therefore, the decisions we make now will dirty the slate again, probably for many decades.

The paper is structured as follows. \S\,\ref{sec:ccditr_terms} follows with definitions of terms, variables and assumptions. \S\,\ref{sec:ccditr_tensions} states a number of scaling requirements that are mutually in tension, then \S\,\ref{sec:ccditr_comp} proposes various ways to resolve these tensions. Some unequivocally solve the dilemmas, others are compromises that partially satisfy some of the apparently mutually incompatible requirements.

\subsection{Terminology \& Assumptions}\label{sec:ccditr_terms}

\begin{assumption}\label{ass:bneck}
Scaling of topology is not part of this exercise. For traffic scaling purposes, it will be sufficient to consider a mini-scenario of a number of flows competing for capacity at a single bottleneck. 
\end{assumption}

Consider a bottleneck link of capacity \(X\) serving a number of traffic flows indexed by \(i\), where each flow has:
\begin{itemize}[nosep]
	\item bit-rate \(x_i\);
	\item round trip time \(R_i\); 
	\item and consists of segments of typical (usually maximum) size \(s_i\). 
\end{itemize}

The number of segments sent but not acknowledged by source \(i\) is termed its window, 
\begin{equation}
W_i = \frac{x_i R_i}{s_i}.\label{eqn:W}
\end{equation}

\begin{assumption}\label{ass:fifo}
We initially assume first-in first-out (FIFO) queuing, so all L4S microflows for a site (customer/user) will share the same queuing delay, \(q\).
\end{assumption}

The round-trip time \(R_i\) of flow \(i\) consists of the base propagation delay between the endpoints \(R_{0i}\) and the queuing delay, that is \(R_i = R_{0i} + q\)

It is unlikely that carrier-scale equipment will implement per-microflow queuing, not only due to cost, but also due to concerns over the tension between transport layer packet inspection and network layer encryption for privacy. Also per-flow queuing in the network requires the network to schedule each microflow, which raises concerns over constraining application flexibility (e.g. variable-bit-rate video).

\begin{assumption}\label{ass:prob_marking}
We assume an L4S-enabled bottleneck implements some form of unary per packet explicit congestion notification (not necessarily the standardized form of ECN~\cite{IETF_RFC3168:ECN_IP_TCP}) so all flows share the same packet marking probability \(p\), with \(0 \le p \le 1\). 
\end{assumption}

Assumption~\ref{ass:prob_marking} does not preclude congestion controls that use both delay and explicit marking as complements. It does imply that solutions based solely on delay and/or loss would require a completely different analysis. 

\begin{assumption}\label{ass:tcp-coexist}
We assume traffic will sometimes share the link with legacy (`Classic') TCP traffic, but it will be isolated from the harmful large and variable queue induced by `Classic' TCP using, for example, the DualQ Coupled AQM~\cite{Briscoe15e:DualQ-Coupled-AQM_ID}.
\end{assumption}
\section{Scalable Congestion Control Tensions}\label{sec:ccditr_tensions}
Here we show that a number of ideal scaling requirements are not all mutually compatible:
\begin{enumerate}[nosep]
	\item Scalable congestion signalling;
	\item Limited RTT-dependence;
	\item Unlimited responsiveness;
	\item Low relative queuing delay;
	\item Unsaturated signalling;
	\item Coexistence with Classic TCP.
\end{enumerate}

The scope is limited to scalability under steady-state conditions. Nonetheless, the purpose of some of the requirements (e.g.\ scalable control signalling) is to enable scaling of dynamic control. However, that linkage will not be explored in this paper.

\subsection{Scalable Congestion Signalling}\label{sec:ccditr_scalable-sig}

\begin{requirement}\label{req:scalable-sig}
For all flows, in the steady state, the number of congestion signals per round-trip, \(v_i\) should be no less than a minimum. 
\end{requirement}

Formally:
\begin{equation}
	v_i \triangleq p W_i;\qquad v_i \ge v_0.\label{eqn:scalable-sig}
\end{equation}
where \(v_0\) will be a widely agreed lower bound for all flows. \(v_0\) need not be \(>1\), but it should not be a lot less than 1, so that even in the worst case (steady-state) there will still be a signal nearly every round trip. The dependence of the variable \(v_i\) on other variables will be investigated below.

This requirement ensures that flow rate can hug variations in available capacity as tightly as possible within the minimum delay that feedback takes to reach the sender. It ensures adjustments can remain as small as possible, which minimizes excursions into both queuing delay and under-utilization.

It also ensures that the sender can detect the absence of congestion signals within a small number of round trips, which can be used to rapidly trigger probing for more available capacity.

\subsection{Limited RTT-Dependence}\label{sec:ccditr_ltd-rtt-dep}

\begin{requirement}\label{req:ltd-rtt-dep}
In the steady state, the throughput of a flow with very large base RTT should not approach starvation while other flows sharing the same bottleneck queue receive plenty.
\end{requirement}

This requirement is deliberately not as strong as `RTT-Fairness', in which the bit rates of competing flows are  required to be independent of their RTTs. Nonetheless, any dependence of flow rate on RTT is required to be limited. The word `fairness' is deliberately not used for this requirement, because it is not trying to describe a desire for different users to have similar rates. Rather it describes the desire that no flow should approach starvation while other flows do not~\cite{Floyd07:Simple_BE}.

It may be argued that existing congestion controls are RTT-dependent, and the lower throughput of large-RTT flows has not been problematic. However, this is because the RTT-dependence of TCP has been cushioned by queuing delay, which L4S aims to remove.

Specifically, with tail-drop queues, the RTTs of all long-running flows have included a common
queuing delay component that is no less than the worst-case base RTT (due
to the historical rule of thumb for sizing access link buffers at 1 worst-case
RTT). So, even where the ratio between base delays is extreme, the ratio
between total RTTs rarely exceeds 2 (e.g.\ if worst-case base RTT is 200\,ms,
worst-case total RTT imbalance tends to \((200 + 200)/(0 +200)\).

Classic AQMs reduce queuing delay to a typical, rather than worst-case,
RTT, but this still cushions the effect of RTT-dependence. For instance, with PIE, the target queuing delay common to each flow is 15\,ms. Therefore, even if the ratio between RTTs is 100\(\times\) (e.g. 200\,ms\(/2\,\)ms) worst-case rate imbalance is only roughly 13 (see \autoref{tab:RTT-imbal}).

However, because L4S all-but eliminates queuing
delay, any RTT-dependence translates (nearly) directly into rate imbalance.
For instance, if the target L4S queuing delay is 500\,\(\mu\)s, the same 100\(\times\) imbalance of base RTTs leads to a rate imbalance of about 80 (also shown in \autoref{tab:RTT-imbal}).

\begin{table}
\centering
\begin{tabular}{p{0.1\columnwidth}rrl}
    \hline
	             & \(q\)~~~~                  & \multicolumn{2}{c}{Total RTT Imbalance}\\\hline
	Drop Tail & 200\,ms        & \((200+200) / (2+200)\) &\(\approx 2\) \\
	PIE AQM & 15\,ms                & \((200+15) / (2+15)\)     &\(\approx 13\) \\
	L4S AQM & 500\,\(\mu\)s   & \((200+0.5) / (2+0.5)\)   &\(\approx 80\) \\\hline
\end{tabular}
\caption{Cushioning effect of queuing delay, \(q\) on Total RTT Imbalance \((R_1 + q) / (R_2 + q)\) for example base RTTs \(R_1=200\)\,ms, \(R_2=2\)\,ms}\label{tab:RTT-imbal}
\end{table}

It is hard to state requirement~\ref{req:ltd-rtt-dep} precisely. Initially it will be stated as a rough equality requirement, but this will be nuanced in later discussion. For now, consider two flows \(i\) \& \(j\) sharing the same bottleneck. Then,
\begin{equation}
	x_i \approx x_j.\label{eqn:rtt-indep-interim}
\end{equation}

Combining \autoref{eqn:W} \& \autoref{eqn:scalable-sig}, the marking probability \(p\) is common to all flows. That is,
\begin{equation*}
	p = \frac{v_i s_i}{x_i R_i} = \frac{v_j s_j}{x_j R_j}.
\end{equation*}
Substituting into \autoref{eqn:rtt-indep-interim}:
\begin{equation}
	\frac{v_i s_i}{v_j s_j} \approx \frac{R_i}{R_j}.\label{eqn:rtt-indep}
\end{equation}	

Therefore, for flow bit-rate to be (roughly) independent of RTT, source \(i\) would have to make either the segment size \(s_i\) or \(v_i\) (or the product of both) proportionate to its RTT \(R_i\). Both lead to problems when the RTT is small:
\begin{itemize}
	\item The segment size is usually set to the maximum that all the links along the path can support. Therefore, to make \(s_i\) proportionate to \(R_i\), segment size would have to be reduced on shorter RTT paths. Then the packet processing rate and therefore the likelihood of processor overload, would be much higher than necessary whenever content was sourced locally, rather than remotely. Such perverse inefficiency is not a feasible proposition.
	\item If \(v_i\) were proportionate to \(R_i\), the number of round-trips between signals would become very large over short RTT paths, leading to slack control of dynamics (failing Req \ref{req:scalable-sig});
\end{itemize}

We will tease apart this dilemma between requirements \ref{req:scalable-sig} \& \ref{req:ltd-rtt-dep} when we consider potential compromises between requirements in \S\,\ref{sec:ccditr_comp}.

One possible escape from this dilemma is that the range of feasible RTTs will not need to scale infinitely, although this point is controversial:
\begin{itemize}
	\item The RTT in glass over the earth's surface between two points at opposite poles (200\,ms or 240\,ms allowing for typical indirect routing) could be considered as an upper bound to RTT. However, this excludes inter-planetary communication, which is likely to become less and less unusual.
	\item There is clearly a minimum distance and therefore RTT between two machines capable of running application processes and congestion control algorithms, given physics sets a minimum bound on the size of transistors. But such a limit would be hard to pin down precisely.
\end{itemize}
Traditionally, flows at very different scales of RTT do not coexist in the same bottleneck. Instead, a domain at one scale (e.g.\ a data centre) is often separated from a domain at another scale (e.g.\ the public Internet) by an intermediate buffering node; a congestion control proxy. Nonetheless, one purpose of designing scalable control algorithms is to remove the need for such proxies.

\subsection{Unlimited Responsiveness}\label{sec:ccditr_unltd-response}

\begin{requirement}\label{req:unltd-response}
An L4S congestion controller must continue to remain responsive to congestion for all values of the window,  \(W_i\).
\end{requirement}

The ACK-clocking mechanism of Classic TCP cannot work if the window is less than \(d\) segments, where \(d\) is the delayed ACK factor. For example, with a delayed ACK factor of 2, the ACK-clock fails if the window is less than 2. If the base RTT is so low that the window needs to be below \(d\) to fit available capacity, Classic TCP never reduces its congestion window below \(d\). Instead, TCP holds the congestion window at \(d\), which forces the queue to grow. This grows the total RTT until a window of \(d\) packets will fit within it~\cite{Briscoe15c:TCP-sub-mss-w}. 

Traditionally, it was thought that this was only a problem with very low capacity. However, once queuing delay is all-but removed, it is not uncommon for the base RTT to be low enough to exhibit this problem. For instance, consider available capacity \(x_i=2\)\,Mb/s, which might occur when a few flows happen to be sharing the link. With a common segment size \(s_i=12\)\,kb and base RTT \(R_i=6\)\,ms, the window to fill this capacity is \(W_i=x_iR_i/s_i=1\) segment per round. 

If L4S controllers became unresponsive at some limit, like Classic TCP does, they would ruin the low queuing delay feature of the L4S service in many realistic cases like those above. This is why requirement \ref{req:unltd-response} states that an L4S controller must not exhibit such a limit.

\subsection{Low Relative Queuing Delay}\label{sec:ccditr_q-rel-low}

\begin{requirement}\label{req:q-rel-low}
Queuing delay should remain small relative to the likely shortest base RTT of any bottlenecked flow.
\end{requirement}

There is no need for queuing delay to be smaller than some absolute delay limit as long as it does not have significant impact relative to the base round trip delay of any communication. 

Queuing delay has been steadily eroded as we have moved from i) tail drop to ii) AQMs for Classic TCP traffic to iii) an L4S AQM designed to be used with Scalable congestion controls. Assuming the bottleneck is in a link where flow multiplexing is low, these respectively keep queuing delay to i) the worst-case base RTT; ii) a typical base RTT; and iii) the minimum expected base RTT. Therefore L4S brings us to the point where this requirement can be satisfied. 

Each case is briefly explained in the following:
\begin{enumerate}
	\item In a low stat-mux case, a well-sized drop-tail buffer is not configured smaller than 1 worst-case bandwidth-delay product, which equates to 1 worst-case RTT of delay. Otherwise all lone flows except those with worst-case RTT would under-utilize the bottleneck link and continual unavoidable bursts would exacerbate under-utilization even for the longest RTT flows.

	\item An AQM is designed to absorb bursts up to a worst-case RTT in duration, so it can be configured to aim for a typical RTT of queuing delay, accepting that there will be some under-utilization by lone large RTT flows. For instance, an AQM in a data centre is configured with a much lower target delay than an AQM in the public Internet. 

	\item The utilization of Scalable traffic is relatively insensitive to a lower-than-optimal target delay~\cite{Alizadeh11:DCTCP_Analysis} so an L4S queue can be configured for close to the minimum likely RTT with very little under-utilization.
\end{enumerate}

If we aim to enable a wider range of flows to coexist in the same bottleneck (e.g.\ 1\,\(\mu\)s--200\,ms), it will be necessary to either manually configure target delay lower, to reflect the lowest typical base RTT, or perhaps to design AQMs that auto-detect the lowest RTT flow that is using the bottleneck at any one time and auto-tune its target delay accordingly.

Satisfying this scaling requirement for a wider range of RTTs seems to require a change to AQM algorithms in the network. Whereas the other requirements have so far been addressed with host-only changes. Nonetheless, this requirement is still relevant to mention here because it complements requirements \ref{req:ltd-rtt-dep} \& \ref{req:unltd-response}. 

\subsection{Unsaturated Signalling}\label{sec:ccditr_sig-sat}

\begin{requirement}\label{req:sig-sat}
Algorithms should avoid saturating congestion signalling at 100\% marking.
\end{requirement}

From \autoref{eqn:scalable-sig} and the valid range of \(p\),
\begin{align}
	 p &= \frac{v_i}{W_i};	&p \le 1.\notag\\
\intertext{\autoref{eqn:scalable-sig} is conditional on \(v_i \ge v_0\), therefore}
	v_0 &\le v_i \le W_i,\notag
\intertext{This combination of inequalities implies \(v_0\le W_i\). So, when the window is small, congestion signalling could saturate at \(p=100\%\). Then the controller will effectively stop reducing the window \(W_i\) in response to further increases in congestion, contravening requirement \ref{req:unltd-response}. This causes the queue to grow, until the total RTT grows large enough to satisfy (substituting from \autoref{eqn:W}):}
	R_i &\ge \frac{v_0 s_i}{x_i}.
\end{align}

\begin{figure}[h]
  \centering
  \includegraphics[width=1.05\linewidth]{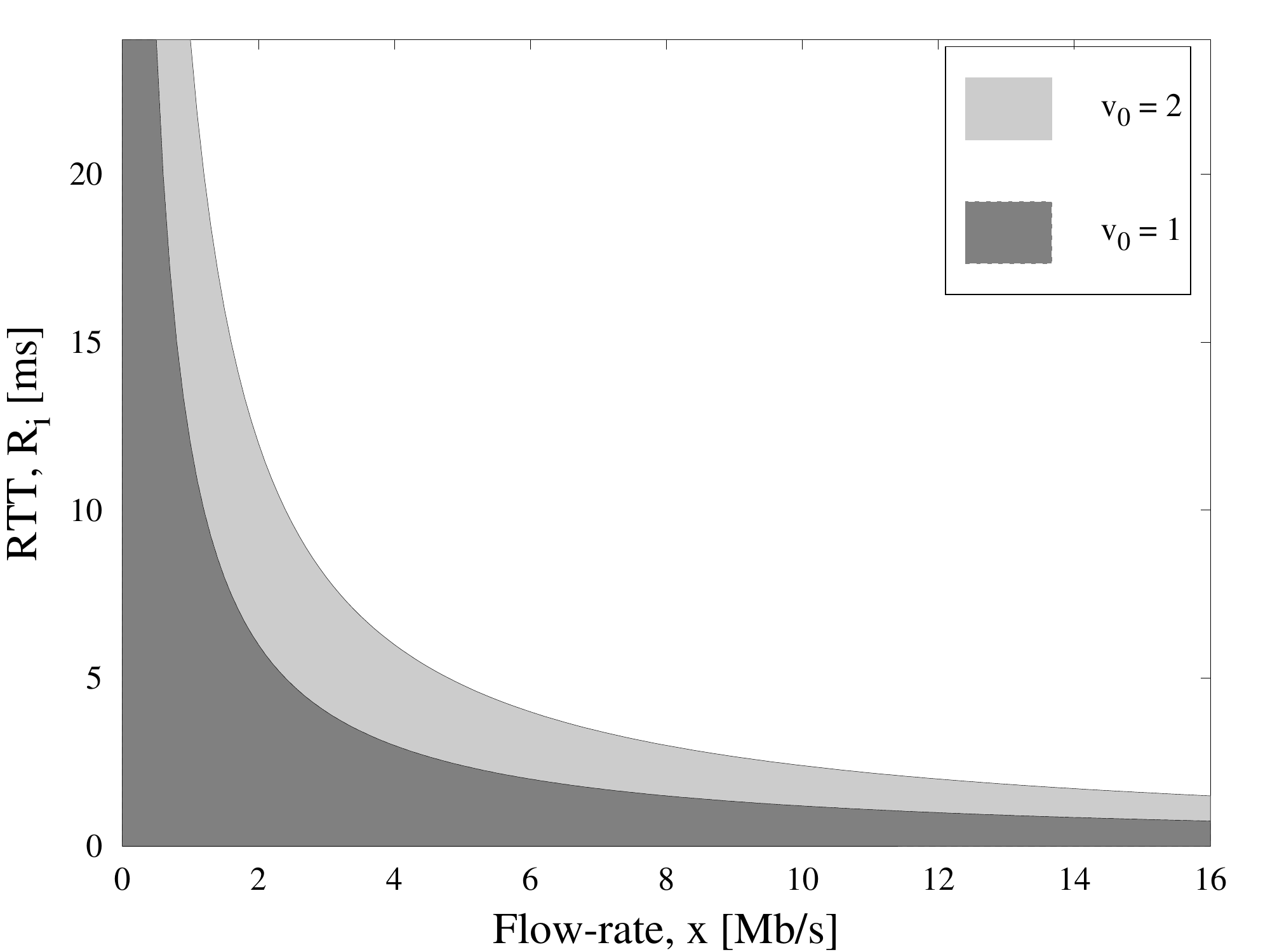}\\
  \caption{Congestion signalling saturates for combinations of RTT and available capacity in the shaded region (segment size, \(s_i=12\,{\mathrm{kb}}\) }\label{fig:R-bound}
\end{figure}

This inequality is plotted in \autoref{fig:R-bound} to illustrate the region where signalling saturates for two example values of \(v_0\).

It might seem that \(v_0\) could be set as low as possible to reduce the likelihood of saturation. However, at the other end of the window spectrum, this would reduce the number of control signals per RTT, compromising  requirement \ref{req:scalable-sig}. 

\subsection{Coexistence with Classic TCP}\label{sec:ccditr_tcp-coexist}

\begin{requirement}\label{req:tcp-coexist}
No standard Classic TCP flow~\cite{IETF_RFC5681:TCP_algorithms} should be pushed towards starvation while any L4S flows are not.
\end{requirement}

As with requirement \ref{req:ltd-rtt-dep}, the words `fairness' or `TCP-friendliness' are deliberately not used for this requirement, because it is not trying to justify some unsubstantiated feeling that different users or applications should have similar rates~\cite{Briscoe06g:Rate_fair_Dis}. It is expressed in terms of each flow avoiding starvation, which Floyd and Allman~\cite{Floyd07:Simple_BE} explained was the underlying motivation behind TCP-fairness. Per-flow starvation-avoidance is all that is necessary for end-systems to implement. Networks might (and often do) additionally enforce or police the relative rates of \emph{users}, but networks need to be careful not to limit application flexibility without strong reasons.

As long as there is plenty of capacity, this requirement then allows flows to weight their rates to be different from each other as long as they do not increase congestion to a level at which a standard TCP flow~\cite{IETF_RFC5681:TCP_algorithms} would approach starvation. 

We use the term `approach starvation' rather than just `starvation' because strictly starvation is a condition where one congestion control continually reduces another, driving it to its minimum throughput whatever capacity is available. It will probably be necessary to define and standardize `approaching starvation' as some minimum throughput of a Classic TCP flow, or equivalently some maximum level of Classic drop (or Classic marking). We shall call these the `tolerable throughput' or `tolerable congestion level', but not quantify them here.

The RTT used by the comparable standard TCP flow also needs to be considered, for two reasons: 
\begin{itemize}
	\item As was explained in \S\,\ref{sec:ccditr_ltd-rtt-dep}, RTT-dependent congestion controls are no longer cushioned by queuing delay when queuing delay is kept low by AQMs. 
	\item The traditional definition of TCP-fairness has always applied to flows of similar RTT, but this is not appropriate for comparing flows that are served by queues with different target queuing delay within the same bottleneck (as in the DualQ AQM~\cite{Briscoe15e:DualQ-Coupled-AQM_ID}). 
\end{itemize}
	
It would be over-restrictive to prohibit Scalable flows from pushing a long RTT Classic flow towards starvation, given short RTT Classic flows already push long RTT Classic flows towards starvation. 

We take the position that we only have to prohibit Scalable flows from pushing low-RTT Classic flows towards starvation. Just as the aim here is to design Scalable CCs with limited RTT-dependence (requirement \ref{req:ltd-rtt-dep}), it can be asserted that there is nothing to stop Classic CCs being redesigned for limited RTT-dependence. If so, there is no doubt that the aggression of long-RTT Classic flows would be increased, rather than that of short-RTT flows decreased. 

This still begs the question of what RTT we mean by a `low-RTT' Classic flow. The RTT of a Classic TCP flow will never be less than the queue delay target in an AQM for Classic traffic. As explained in \S\,\ref{sec:ccditr_q-rel-low} the queuing delay target of a Classic AQM is configured for the typical RTT of the flows it controls. We do not know of a study that measures the average base RTT of traffic on the pubic Internet weighted by usage. Nonetheless, the lowest opinion of what is `typical' is the 5\,ms target of CoDel.\footnote{In private networks, e.g.\ data centres, the typical RTT and therefore the queuing delay target of an AQM will generally be lower.}

By Assumption \ref{ass:tcp-coexist}, L4S and Classic traffic share capacity through a mechanism like the DualQ Coupled AQM~\cite{Briscoe15e:DualQ-Coupled-AQM_ID}. Currently, this relates the loss (or ECN marking) probability seen by Classic traffic, \(p_C\), to that seen by L4S traffic, \(p\), as follows:
\begin{align}
	p_C = \left(\frac{p}{k}\right)^2.\label{eqn:dualq-coupled}
\end{align}
By the above arguments, it is sufficient for coexistence to set the coupling factor, \(k\) with only `low-RTT' Classic flows in mind. 

This leaves the question open of what value to agree on for the aggressiveness of Scalable flows (\(v_0\) in \autoref{eqn:scalable-sig}). The choice of \(v_0\) already requires a tough compromise to be struck between requirements \ref{req:scalable-sig}, \ref{req:ltd-rtt-dep}, \ref{req:unltd-response} and \ref{req:sig-sat}. So it would be sensible to wait for some consensus to emerge over the choice of \(v_0\) before recommending a value for the coupling factor \(k\).

\section{Solutions and Compromises}\label{sec:ccditr_comp}

\subsection{Unsaturated Marking}\label{sec:unsat}

A scheme such as REM~\cite{Athuraliya01:REM} could be used in the network to reduce the likelihood that signalling will saturate (requirement \ref{sec:ccditr_sig-sat}). Nonetheless, below we propose a scheme that purely involves the sender's control algorithm.

It is proposed to use the number of unmarked packets, \(u\), between marked packets to drive the sender's congestion control algorithm. If \(p\) is the packet marking probability, as already defined, then the number of packets delivered per marked packet is \(1/p\).  Therefore, the number of unmarked packets between the marked packets,
\begin{align}
	u &= \frac{1}{p} - 1\notag
\intertext{or}
	\frac{1}{u} &= \frac{p}{(1-p)}\label{eqn:unsat}
\end{align}
Whereas \(p\) is confined to the range \([0,1]\), the range of \(1/u\) is \([0, \infty)\). This is the unsaturating property that is needed. 

\begin{figure}[h]
  \centering
  \includegraphics[width=\columnwidth]{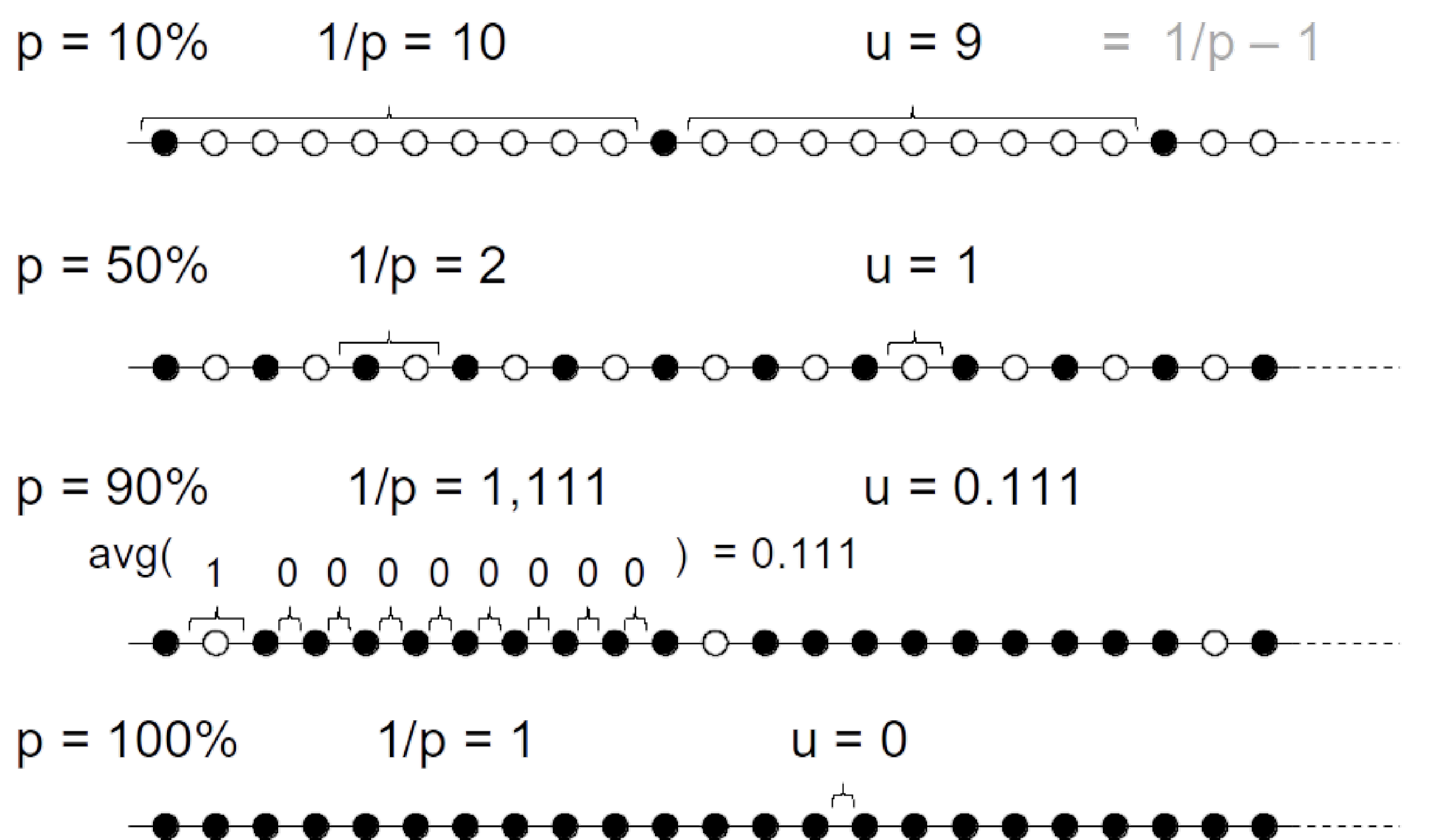}\\
  \caption{Number of unmarked packets as a non-saturating congestion signal}\label{fig:unsat}
\end{figure}

Under normal lightly loaded conditions, when \(p\) is close to zero, \(1/u\rightarrow p\). For example, if \(p=0.01,  1/u=0.01010101\). But as \(p\rightarrow1, 1/u\rightarrow\infty\). For example, if \(p=0.9999999, 1/u=9999999\). Other examples are illustrated in \autoref{fig:unsat}.

These unsaturating congestion signals will sometimes be called virtual marks, because the host (or any observer) can calculate the occurence of virtual marks from the spacing between real marks.

\subsection{Scalable Signalling vs.\ RTT Independence}\label{sec:Sig_v_RTT}

A difficult tension remains between the scalable congestion signalling requirement (\ref{sec:ccditr_scalable-sig}) and the requirement to limit RTT-dependence (\ref{sec:ccditr_ltd-rtt-dep}). 

The authors cannot find an elegant resolution to this tension. Instead, we have considered a number of inelegant compromises. Those ideas that are decent enough to present here are named ``Compromise 4'' and ``Compromise 5''. The flow of the argument continues from \S\,\ref{sec:ccditr_ltd-rtt-dep}, where we initially set the bit rates of two competing flows, \(i\) \& \(j\) with different RTTs to be roughly equal, \(x_i\approx x_j\), which we shall call our `interim RTT-independence requirement'. Different compromises soften this requirement in different ways.

Before continuing, we shall simplify. \S\,\ref{sec:ccditr_ltd-rtt-dep} concluded that the maximum segment size could not be varied upwards, given existing link limitations, and nothing would be gained by varying it downward. Therefore we shall simplify by discussing packet rate, \(r\), not bit rate. Then \autoref{eqn:W} can be restated as
\begin{equation}
W_i = r_i R_i.\label{eqn:W-wrt-r}
\end{equation}

\subsubsection{Compromise 4}\label{sec:compromise4}

This compromise ends up not being chosen, but it is included here to illustrate the problem.

Returning to our interim RTT-independence requirement, it led us to \autoref{eqn:rtt-indep}, which required that the marks per RTT, \(v_i \propto R_i \). Substituting in \autoref{eqn:scalable-sig},
\begin{align}
	pW_i &\propto R_i\label{eqn:r-rtt-indep-base}
\intertext{Substituting from \autoref{eqn:W-wrt-r} and introducing a constant of proportionality}
	pr_iR_i &= c_0R_i\notag\\
	r_i        &= \frac{c_0}{p}.\label{eqn:r-rtt-indep}
\end{align}

The constant \(c_0=pr_i\) can be interpreted as the constant number of marks per unit time necessary for RTT-independence. For example, if \(c_0=1000\), in each flow, one packet would be marked per ms.

Such RTT-independence would be problematic in two cases:
\begin{description}
	\item[Low rate:] If \(r<1\)\,packet/ms, there would not be enough packets to be marked once per ms;
	\item[Low RTT:] If \(R_i<1\)\,ms, there would be less than 1 mark per round trip. For example, if \(R_i=1\,\mu\)s, there would be only one mark every 1,000 round trips, which would not provide the tight control demanded by requirement \ref{eqn:scalable-sig}.
\end{description}
The first problem is a signalling saturation problem, which can be solved using the technique in \S\,\ref{sec:unsat}. The second problem is not surprising, because \autoref{eqn:r-rtt-indep} is derived from \(pW_i = c_0R_i\), and when \(R_i\) is small this contravenes \(v_i\ge v_0\) from \autoref{eqn:scalable-sig}, which expresses the scalable signalling requirement.

In contrast, DCTCP~\cite{Alizadeh10:DCTCP} is a good example of the advantage of scalable congestion signalling (requirement \ref{eqn:scalable-sig}). In the steady state its congestion window converges to,
\begin{align}
	W_i &= \frac{v_0}{p},\notag
\intertext{where in DCTCP's case \(v_0=2\) segments. However, this contravenes requirement \ref{req:ltd-rtt-dep}, because, substituting from \autoref{eqn:W-wrt-r}, the packet rate is inversely dependent on RTT.}
	r_i &= \frac{v_0}{R_ip}.\label{r-dctcp}
\end{align}

One possible compromise is to replace the dependence on RTT in \autoref{r-dctcp} with dependence on another scalable property, perhaps the inter-packet departure time, \(1/r\):
\begin{align*}
	r_i &= \frac{v_0}{p/r_i}.
\intertext{This should scale reasonably well, because it will only be less than the RTT if the window is less than 1 segment, which is uncommon (but not impossible---see requirement \ref{req:unltd-response}). However, this simplifies to}
	p  &= v_0,
\end{align*}
which would be an impractical rate control, because, the congestion level would not change with rate, so it would never converge.

A possible alternative compromise would be to replace dependence on RTT with dependence on the square-root of the inter-departure time \(1/sqrt{r}\). For completeness, we will also address the saturation problem by replacing \(p\) with \(1/u\):
\begin{align}
	r_i &= \frac{v_0u}{1/\sqrt{r_i}},\notag
\intertext{which simplifies to}
	r_i &= v_0^2u^2.\notag
\intertext{Renaming the squared constant, we get}
	r_i &= c_0u^2.\label{eqn:comp4}
\end{align}

\begin{figure}[h]
  \centering
  \includegraphics[width=1.05\columnwidth]{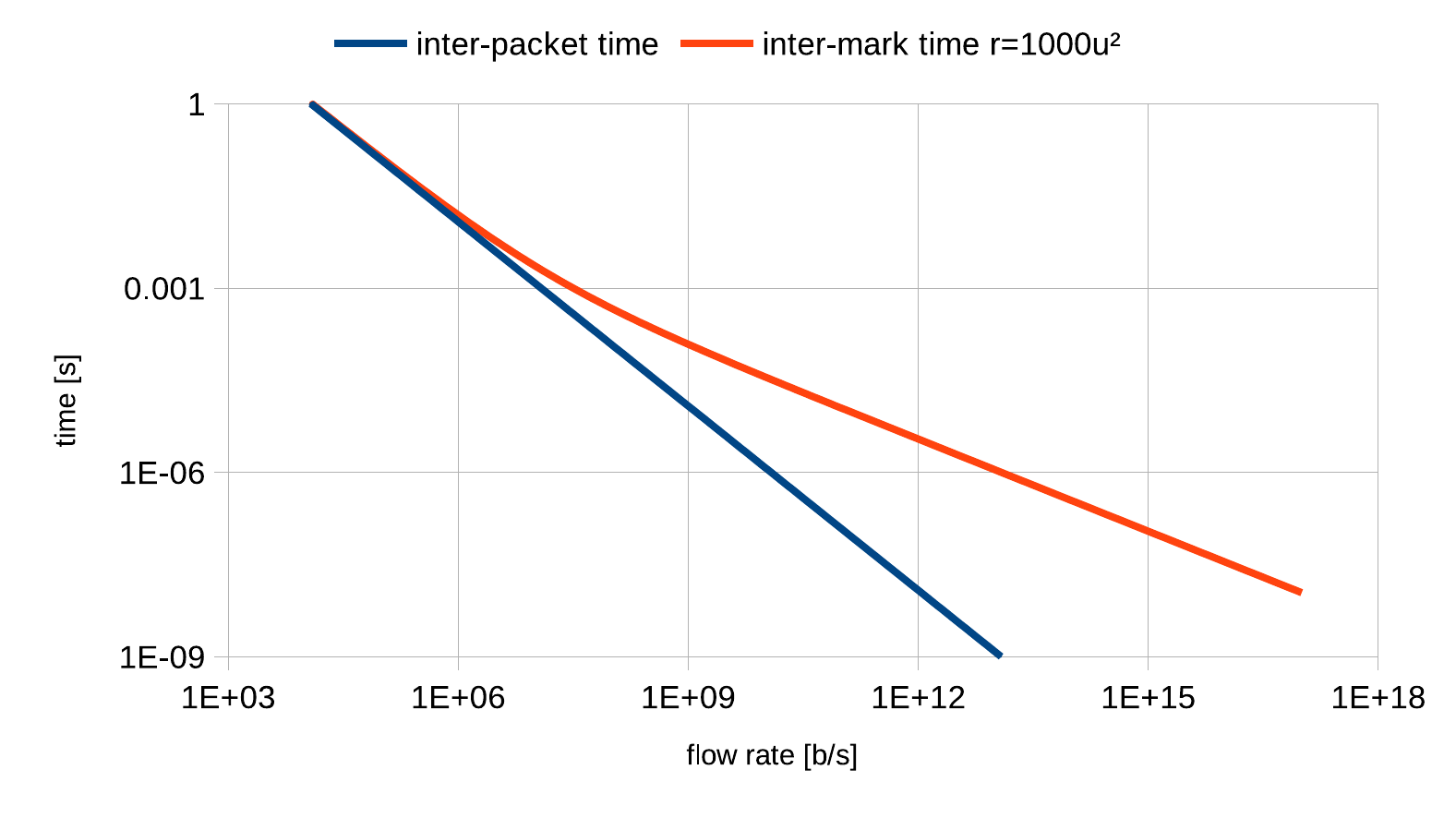}\\
  \caption{Inter-mark time compared with inter-packet departure time for the `Compromise4' algorithm of \autoref{eqn:comp4}}\label{fig:compromise4}
\end{figure}

\autoref{fig:compromise4} uses \autoref{eqn:comp4} to plot the inter-mark time \(1/(pr_i)\) compared to the inter-packet departure time, using \(c_0=1000\). At any flow rate, for example the vertical at 1\,Gb/s, the ratio of the times where this vertical intersects the two plots (120\,\(\mu\)s / 12\,\(\mu\)s = 10) represents a likely worst-case number of round trips per mark at that flow-rate. This assumes that the worst-case is a window of one segment, so that the intersection of the vertical with the inter-packet time plot represents a worst-case RTT, which is perhaps reasonable, but not strictly true, as already discussed.

Therefore, at flow rates below about 100\,Mb/s, there is little likelihood of unscalable control signalling (many round trips between marks). However, at higher flow rates, and low RTTs, this approach compromises the scalable control signalling requirement (\ref{req:scalable-sig}) in favour of RTT-independence.

Further, because of the squared congestion metric in \autoref{eqn:comp4}, the coupling between Classic and L4S congestion signals would have to be altered from that given in \autoref{eqn:dualq-coupled}. In order to coexist with Classic TCP (requirement \ref{req:tcp-coexist}), the coupling would require an exponent of 4, rather than 2.

It is questionable whether it will be worthwhile to standardize an exponent of 4 rather than 2 in the L4S coupling mechanism, solely to support an approach that does not reliably satisfy one of the conflicting requirements, specifically scalable signalling (requirement \ref{req:scalable-sig}). 

\begin{table*}
\centering
\begin{tabular}{lp{0.35\linewidth}p{0.5\linewidth}}
    \hline
	1. & Scalable congestion signalling & Good compromise \#5 (\S\,\ref{sec:compromise5}) or \#4?\\
	2. & Limited RTT-dependence		  & Good compromise \#5 (\S\,\ref{sec:compromise5}) or \#4?\\
	3. & Unlimited responsiveness         & To be resolved\\
	4. & Low relative queuing delay	      & Separate scope: AQM requirement\\
	5. & Unsaturated signalling               & Resolved (\S\,\ref{sec:unsat})\\
	6. & Coexistence with Classic TCP.  & Resolved~\cite{Briscoe15e:DualQ-Coupled-AQM_ID} \\\hline
\end{tabular}
\caption{Status of Steady-State Scaling Requirements}\label{tab:status}
\end{table*}

\subsubsection{Compromise 5}\label{sec:compromise5}

The nub of the tension can be seen by restating the equations representing the scalable signalling requirement (\autoref{eqn:scalable-sig}) and the limited RTT-dependence requirement (\autoref{eqn:r-rtt-indep-base}) together:
\begin{align}
	pW_i &\ge v_0, & (\ref{eqn:scalable-sig})\notag\\
	pW_i &\propto R_i, & (\ref{eqn:r-rtt-indep-base})\notag
\end{align}

\begin{figure}[h]
  \centering
  \hspace{-12pt}
  \includegraphics[width=1.0\columnwidth]{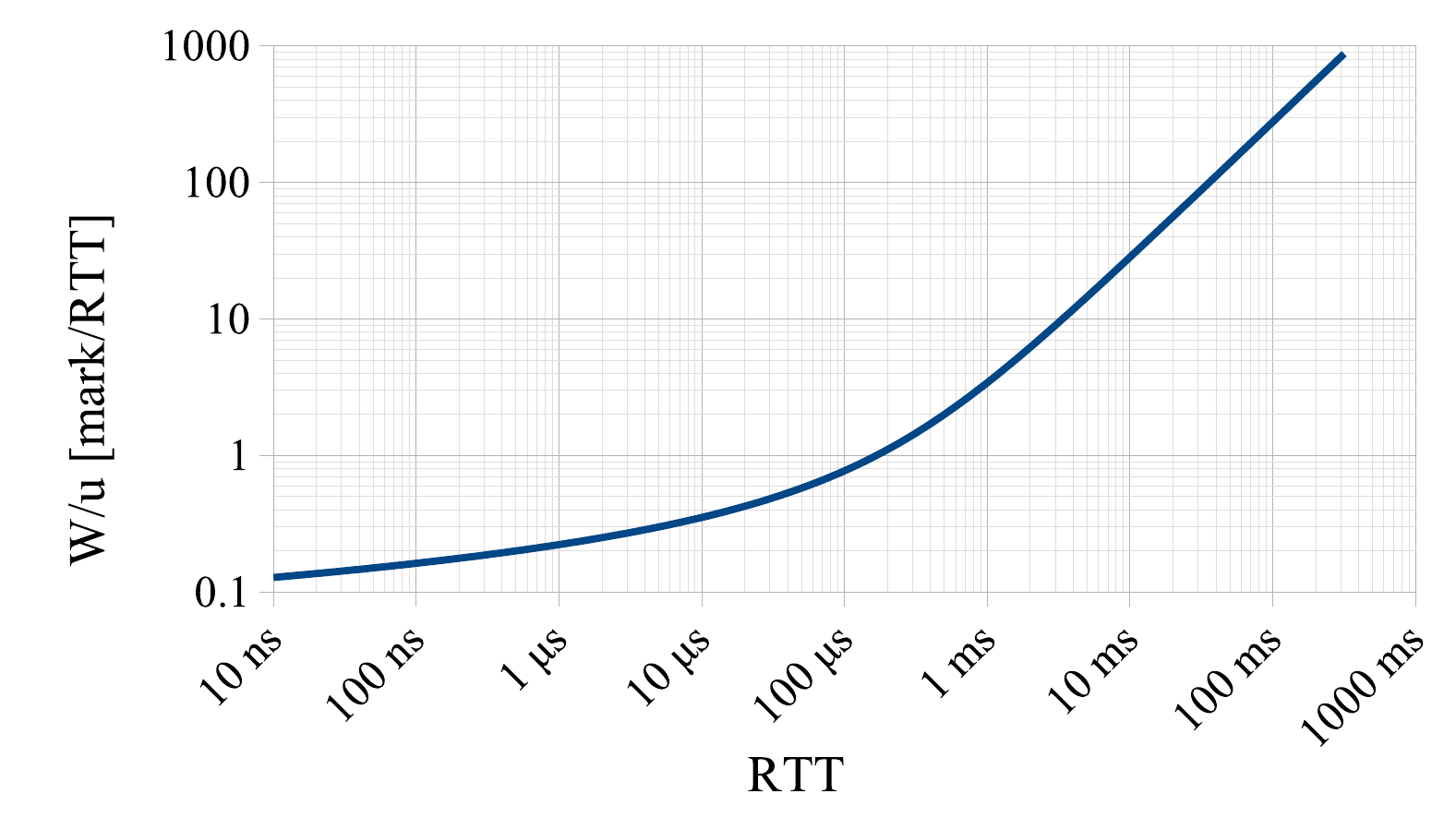}\\
  \caption{Compromise 5 control algorithm showing marks per RTT as a function of RTT (\autoref{eqn:compromise5-W})}\label{fig:compromise5-vmark_RTT}
\end{figure}

A better compromise might be possible if the marks per RTT can take the form of a function of RTT \(v_i(R_i)\), such that, as RTT reduces, marks per RTT are lower bounded (or at least reduce slowly) while, as RTT rises, marks per RTT become proportional to RTT. \autoref{eqn:compromise5-W} fits this description fairly well:
\begin{align}
	\frac{W_i}{u} &= \frac{v_0}{\lg{(R_0/R_i + 1)}}.\label{eqn:compromise5-W}
\end{align}
\(R_0\) would probably need to be standardized, at least to within a range. It is a configuration parameter common to all flows that represents the RTT at which \(W_i/u_i=v_0\). This formula is illustrated in \autoref{fig:compromise5-vmark_RTT} using parameters \(v_0=2, R_0=500\,\mu\)s.

It will be noted that non-saturating congestion signals, \(1/u\), have been used in place of \(p\), as described in \S,\ref{sec:unsat}. We use the unit `marked packet' for these signals, which is a good enough approximation at the low marking probabilities used in the examples here. 

\begin{figure}[h]
  \centering
  \hspace{-12pt}
  \includegraphics[width=1.0\columnwidth]{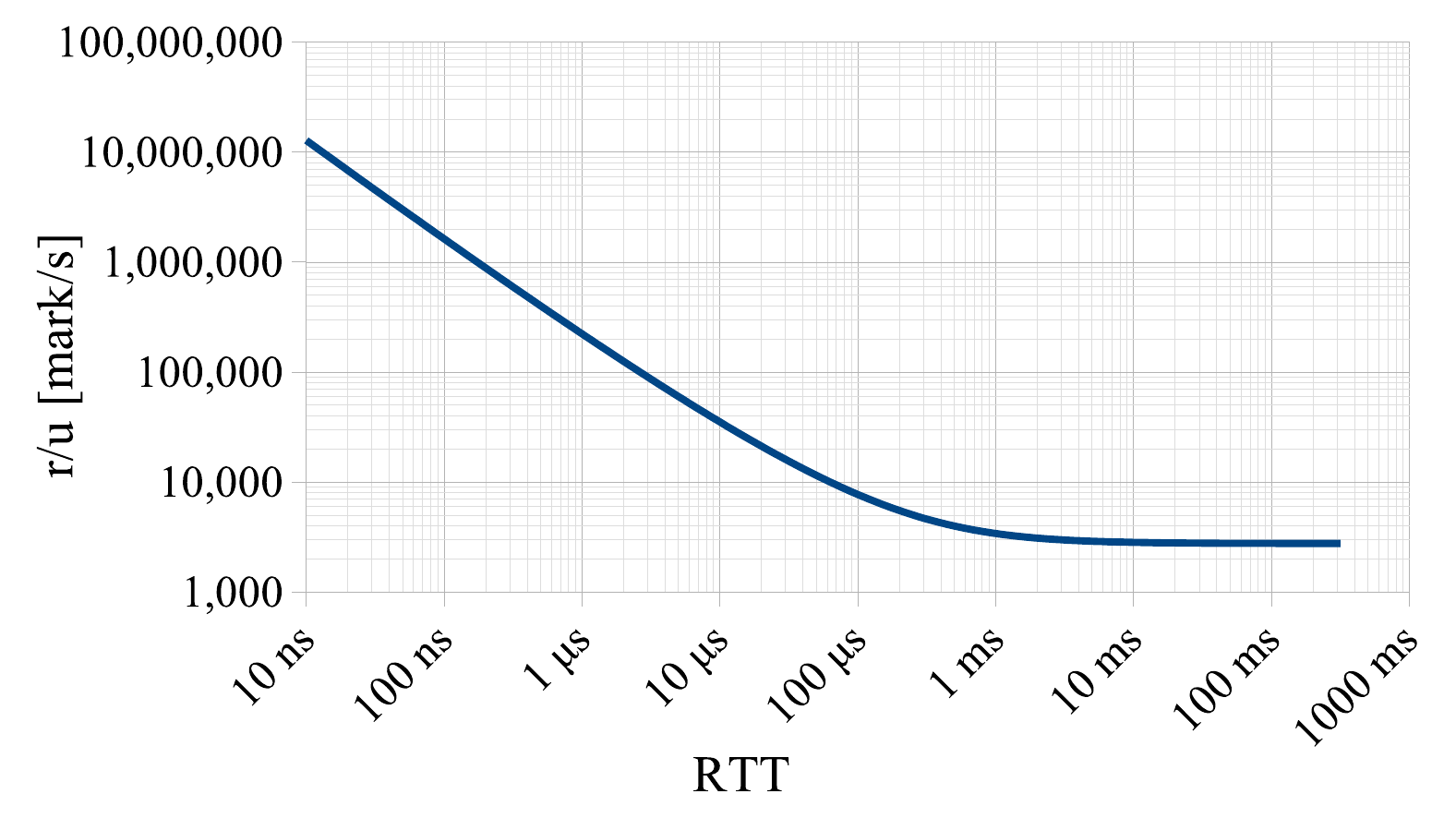}\\
  \caption{Compromise 5 control algorithm showing marks per second as a function of RTT (\autoref{eqn:compromise5-r})}\label{fig:compromise5-vmark_time}
\end{figure}

The marks have been contrived to become proportional to RTT as RTT rises\footnote{The Taylor series of \(\ln(1+y) = y - y^2/2 + y^3/3 \ldots\).\\So as \(y\rightarrow0;\quad\ln(1+y)\rightarrow y\).\\So, for \(R_i \gg R_0; \quad v_0/\lg{(R_0/R_i + 1)}\rightarrow \ln{(2)}v_0 R_i/R_0\)} so that, when marks/RTT is divided by \(R_i\) to derive the formula for marks per second, it will tend to a constant asymptote. The resulting formula for marks per second is given in \autoref{eqn:compromise5-r} and \autoref{fig:compromise5-vmark_time} implies that it does indeed tend to a constant of about 2,800 as \(R_i\rightarrow\infty\).
\begin{align}
	\frac{r_i}{u} &= \frac{v_0}{R_i \lg{(R_0/R_i + 1)}}\label{eqn:compromise5-r}\\
	          &= f(R_i).\label{eqn:compromise5-r-f}
\end{align}

For two flows, \(i\) \& \(j\) with RTTs \(R_i\) \& \(R_j\), the ratio between their packet rates will be the ratio of the functions \(f(R_i)/f(R_j)\) using \autoref{eqn:compromise5-r-f}. This is because \(u\) will always be common to both flows. For example, reading off from \autoref{fig:compromise5-vmark_time} at \(R_i=10\,\mu\)s \& \(R_j=130\,\)ms, \(r_i/r_j\approx35,000/2,800\approx13\). Thus, a round-trip ratio of over 4 orders of magnitude only results in a rate imbalance of a little more than 1 order of magnitude.

This relatively small rate imbalance is not at the expense of control signal scaling. For instance, in a round trip of \(10\,\mu\)s there are about 0.35 marks (about 3 round trips per mark).

Therefore, in theory at least, `Compromise 5' is a good compromise between scaling requirements that were thought to be mutually incompatible. 
\section{Summary}\label{sec:ccditr_discuss}

The status of the requirements set at the start of this document are summarized in \autoref{tab:status}.

The tension between the first two requirements is resolved fairly well by Compromise 5 (\S\,\ref{sec:compromise5}), but this does not preclude finding a better compromise.

The unlimited responsiveness requirement (\ref{sec:ccditr_unltd-response}) was set aside for the purposes of the present paper because it is not so obviously in tension with any other requirements. It remains to be resolved.

\addcontentsline{toc}{section}{References}

{\footnotesize%
\bibliography{ccdi}}


\onecolumn%
\addcontentsline{toc}{part}{Document history}
\section*{Document history}

\begin{tabular}{|c|c|c|p{3.5in}|}
 \hline
Version &Date &Author &Details of change \\
 \hline\hline
00A                   &24-May-2016 &Bob Briscoe &First Draft\\\hline%
00B                  &25-May-2016  &Bob Briscoe &Extended Introduction, Completed TCP Coexistence, added Discussion section. Plus minor clarifications throughout.\\\hline%
00C                  &26-Mar-2017  &Bob Briscoe &Restructured to defer potential compromise solutions until after all constraints. Added new compromise ideas.\\\hline%
00D &27-Mar-2017     &Bob Briscoe &Nits and added brief discussion section.\\\hline%
00E &07-Jul-2017     &Bob Briscoe &Fixed nits and avoided describing \(1/u\) as virtual marking probability.\\\hline
\metaversion &\metadate     &Bob Briscoe &De-garbled Coexistence text; added RTT-imbalance examples.\\\hline%
\hline%
\end{tabular}

\end{document}